\documentclass[%
 aip, apl,
 amsmath, amssymb,
 reprint,%
]{revtex4-1}

\usepackage{graphicx}
\usepackage{dcolumn}
\usepackage{bm}

\usepackage[utf8]{inputenc}
\usepackage[T1]{fontenc}
\usepackage{mathptmx}
\usepackage{etoolbox}

\makeatletter
\def\@email#1#2{%
 \endgroup
 \patchcmd{\titleblock@produce}
  {\frontmatter@RRAPformat}
  {\frontmatter@RRAPformat{\produce@RRAP{*#1\href{mailto:#2}{#2}}}\frontmatter@RRAPformat}
  {}{}
}%
\makeatother
\begin{document}

\title[draft]{Superconducting Pulse Conserving Logic and Josephson-SRAM}
\author{Quentin Herr}
 \email{quentin.herr@imec-int.com}
 \affiliation{imec USA, Kissimmee FL}
\author{Trent Josephsen}
 \affiliation{imec USA, Kissimmee FL}
\author{Anna Herr}
 \affiliation{imec, Leuven BE}

\date{\today}

\begin{abstract}
Superconducting digital Pulse-Conserving Logic (PCL) and Josephson
SRAM (JSRAM) memory together enable scalable circuits with energy
efficiency 100$\times$ beyond leading-node CMOS. Circuit designs
support high throughput and low latency when implemented in an
advanced fabrication stack with high-critical-current-density
Josephson junctions of 1000\,$\mu$A/$\mu$m$^2$. Pulse-conserving logic
produces one single-flux-quantum output for each input, and includes a
three-input, three-output gate producing logical or3, majority3 and
and3. Gate macros using dual-rail data encoding eliminate inversion
latency and produce efficient implementations of all standard logic
functions. A full adder using 70 Josephson junctions has a carry-out
latency of 5\,ps corresponding to an effective 12 levels of logic at
30\,GHz. JSRAM (Josephson SRAM) memory uses single-flux-quantum
signals throughout an active array to achieve throughput at the same
clock rate as the logic. The unit cell has eight Josephson junctions,
signal propagation latency of 1\,ps, and a footprint of
2\,$\mu$m$^2$. Projected density of JSRAM is 4\,MB/cm$^2$, and
computational density of pulse-conserving logic is on par with leading
node CMOS accounting for power densities and clock rates.
\end{abstract}

\maketitle

Single Flux Quantum (SFQ) integrated circuits have shown
slow-but-steady progress in clock rates, bit-error rates,
interconnects, and energy efficiency, with fundamentals established in
small-scale demonstrations. The first demonstrations emphasized high
data rates, with a binary counter reported at 100\,GHz already in 1982
\cite{hamilton1982100}. Measurements of Bit Error Rate (BER) show good
performance across the SFQ logic families including Rapid-SFQ (RSFQ),
Quantum Flux Parametron (QFP), and Reciprocal Quantum Logic (RQL)
\cite{herr1996error, herr1999temperature, herr2011ultra,
  takeuchi2017measurement, herr2022measurement}. SFQ logic families
operate in the thermal limit, meaning the devices are sized to achieve
the desired bit-error rate based on Johnson noise in the Josephson
junctions (JJs) at the operating temperature, typically LHe at 4.2\,K.
 Despite the similarities at the gate
level, SFQ logic families differ in power distribution and timing
characteristics, with widely divergent scalability and performance.

RSFQ has had success in decimation filters for
mixed-signal applications \cite{mukhanov2004superconductor,
  kashima202164} where scale is modest and bit rates are high, but
scale is limited by a DC current draw of 1\,A per 1,000 gates, and by the
timing uncertainty of free-running pulses. Energy Efficient SFQ
(eSFQ/ERSFQ) \cite{mukhanov2011energy, kirichenko2019ersfq} and Dynamic
SFQ (DSFQ) \cite{rylov2019clockless} variants have these same
limitations. Asynchronous approaches \cite{deng1997data,
  tzimpragos2021temporal} can guarantee functional circuits, but at a
cost to throughput, as a reduced rate of computation is the only
solution to timing uncertainty. Partitioning the circuit onto floating
ground planes and ``recycling'' the current has been proposed to
reduce current draw, but the interconnect overheads are high
\cite{johnson2003differential} and demonstrations limited to 16
partitions with nearest-neighbor interconnect
\cite{filippov2009serially, kaplan2012serial, sano2016reduction}. With
current recycling, passive transmission line interconnect is
unsupported and further scaling is in doubt \cite{semenov2019current}.

QFP logic benefits from the scalable power distribution and timing
stability that comes from AC bias. In fact, the resonant clock
distribution network was first put forward in the context of QFPs
\cite{hosoya1991quantum}. Adiabatic QFPs (aQFPs) have switching energy
of 10\,zJ at a 1\,GHz clock rate\cite{takeuchi2013measurement,
  yamae2019systematic}, but the advantages of the quasi-adiabatic
potential are partially offset by the need for larger junctions, so
the energy dissipation for a given BER \cite{takeuchi2017measurement}
is similar to that of RQL \cite{herr2022measurement}. QFPs are used as
auxiliary circuits such as qubit readout \cite{volkmann2015low,
  grover2020fast}, where clock rates are low. Demonstrations have
advanced to a 4-bit CPU core \cite{ayala2020mana}, but there is a
fundamental limitation in scaling to high performance as the
technology supports only one level of logic per clock cycle,
translating to high latency and a low logic clock rate. QFPs have high
dynamic power dissipation despite low switching energy, as every gate
switches every cycle. QFPs do not support transmission line
interconnect \cite{china2016demonstration}.

\begin{figure*}
\includegraphics[width=7.0in]{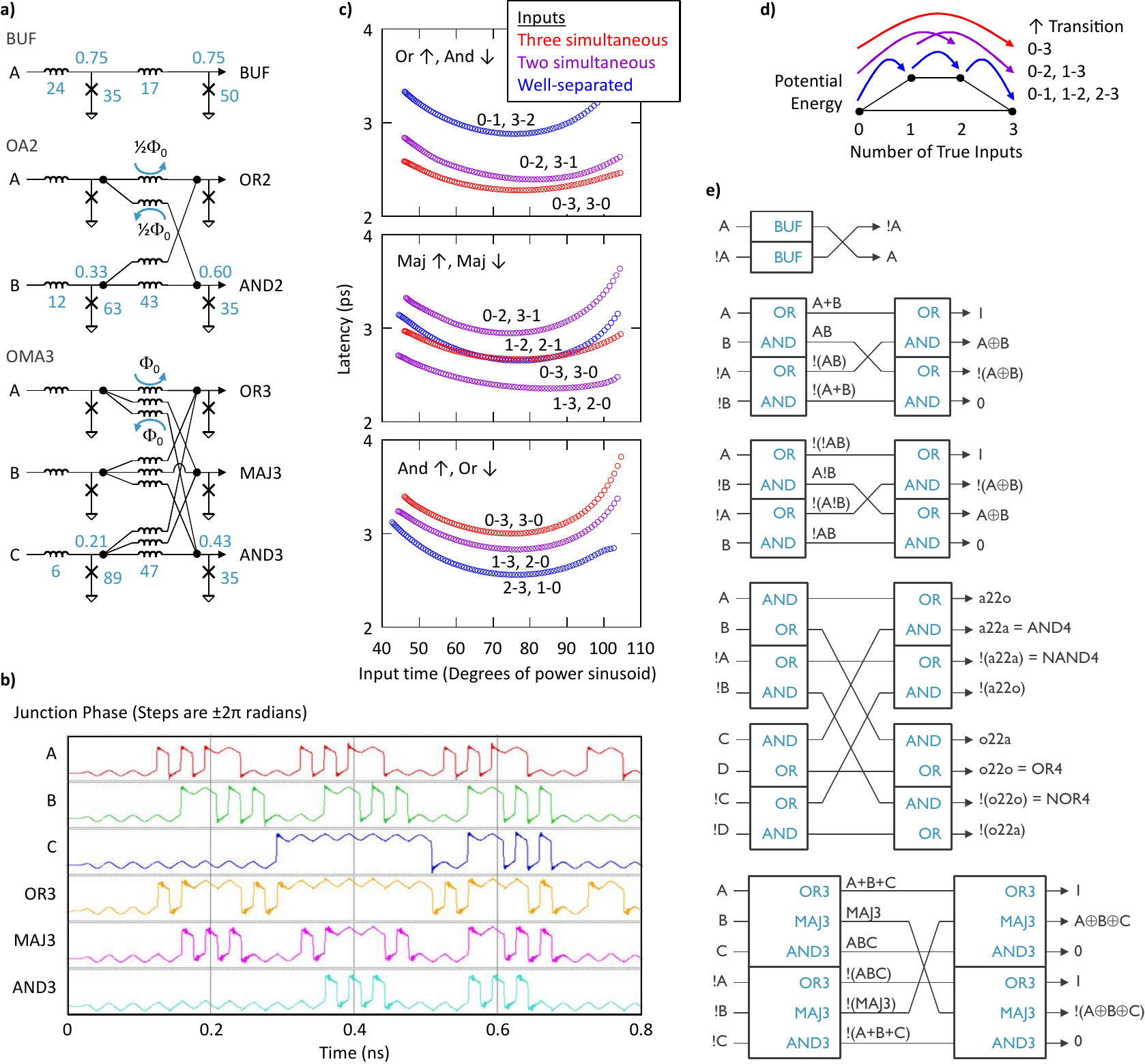}
\caption{\label{fig1} PCL logic schematics, modeling, and macros. a)
  Gate schematics show the single-input JTL buf (buffer) used for
  isolation and gain, the two-input oa2 gate that produces logical or2
  and and2, and the three-input oma3 gate that produces or3, maj3
  (majority), and and3. All junctions are AC biased (not
  shown). Values for inductors are shown in pH, junction critical
  currents in $\mu$A, and peak-amplitude AC bias current as a fraction
  of the junction. b) WRSpice simulation of the oma3 gate showing
  waveforms of junction phase, with each positive SFQ pulse equal to a
  rising edge and each negative pulse equal to a falling edge. c)
  Simulations of oma3 gate delay as a function of input time, using
  1\,mA/$\mu$m$^2$ Si-barrier junctions, shown for different initial
  states and relative input pulse timings. Positive-pulse delays on Or
  correspond to negative-pulse delays on And, and vise versa. d) A
  diagram of the potential energy of SFQ stored within the gate, and
  trajectories between states that correspond to the gate delays in
  (c). For clarity, only positive-pulse transitions are shown. e) PCL
  dual rail macros produce a complete set of standard logic functions
  using only positive-logic primitives.}
\end{figure*}

RQL has stable timing, multiple levels of logic per pipeline stage,
low power dissipation and low BER, and low junction count
\cite{herr2011ultra, herr2022measurement, herr20138}.  Demonstrations
of superconducting transmission line interconnect include synchronous
data links chip-to-chip \cite{egan2022synchronous} and isochronous
links for going board-to-board \cite{talanov2022propagation,
  dai2022isochronous}. Highly functional 8- and 16-bit CPUs have been
presented \cite{vesely2018pipelined} with AC power distribution at
scale \cite{strong2022resonant}.  Because SFQ logics are pulse-based,
inversion is expensive.  The inverting RQL ``AnotB'' gate
\cite{herr2011ultra} has setup time between the inputs, which adds
latency in proportion to the local timing uncertainly despite stable
timing on a global scale. Phase Mode Logic (PML)
\cite{carmean2017phase} encodes logical inversion as a signal-polarity
inversion in analogy to CMOS, but the PML data rate is limited to half
the resonator clock rate, and inversion has half a cycle of latency.

In this letter we present a new Pulse Conserving Logic (PCL) and
Josephson SRAM (JSRAM) memory that have the potential to compete with
conventional CMOS in integration scale and performance. These SFQ
designs fully exploit the resources of the fabrication process
proposed in a companion letter. The fabrication stack and resonant
power network apply equally to logic and memory and support up to 400M
AC-biased JJs per cm$^2$. Fabrication stack resources were codesigned
with the physical layouts of the circuits and carefully balanced to
provide dramatic improvements in density relative to legacy designs.

We have developed a new Pulse Conserving Logic (PCL) that uses dual
rail signals. Like, CMOS, energy is dissipated only on transitions,
and there is no penalty in clock rate or latency for inversion. Our
logic gates have the same number of inputs as outputs and preserve
pulses, meaning each input SFQ is routed to an output. Gate outputs
have symmetric double-well potentials, improving parametric margins
and power dissipation. Conservation of pulses enables complex logical
functions at reduced latency. The primary set of PCL gates is shown in
Fig.~\ref{fig1}a. The Josephson Transmission Line (JTL) is a buffer
that passes the pulse to a single output while providing isolation and
transimpedance gain \cite{likharev1991rsfq}. The oa2 gate passes the
first pulse to an or2 output, and the second pulse to an and2. The
three-input, three-output oma3 gate passes the first pulse to o3, the
second to maj3 (majority function), and the third to and3. Fanout of
the buf is two, and fanout of the oa2 is one-half, meaning a buf stage
is required between interconnected gates. Fanout of the oma3 is
one-quarter, meaning two buf stages are required between gates, with
the second stage sized up by a factor of two.

Compared to an earlier implementation, the oa2 logic gate does not
require transformers between the signal inductors \cite{herr2011ultra}
and does not have inductors between outputs or from an output to
ground. These advantages permit generalization to oma3.  All inputs
connect to all outputs, and the applied flux bias on the interconnect
distributes the output junctions along a Josephson phase wheel
\cite{lee1993phase}. Negative inputs pulses to the gates reverse the
rotation of the phase wheel. Extension to a four input, four-output
gate is thus possible but would require four flux biases and high
drive strength on the inputs. To produce the standard logic functions
such as or4 and and4, a two-level cascade of two-input gates is more
efficient.

As with the other ac powered logics, the SFQ signals are bipolar, with
the positive pulse representing a transition to logical ``1'' and a
negative pulse to logical ``0''. Waveforms of superconducting phase as
shown in Fig.~\ref{fig1}b resemble conventional level-based
logic. Both low and high levels can persist for multiple clock cycles.

Circuit simulations use a Josephson junction model with nominal
critical current, $I_c$, of 100\,$\mu$A, internal capacitance of
3.5\,fH, and voltage gap of 2.75\,mV. The bias tap for a standard JTL
junction is modeled as sinusoidal voltage source in series with
6.2\,fF capacitor. Both capacitors shunt the device, as the bias tap
is connected to signal ground, and both scale with device critical
current. The junctions are self-shunted, meaning the combined shunt
capacitance is low enough that the gap voltage of itself prevents
hysteresis in the IV curve. The junction internal resistance has only
a secondary effect on switching speed. We include an external shunt
resistance, $R$, such that $I_cR=4.4$\,mV this allows voltage sub-gap
plasmon oscillations to ring down between SFQ switching events.

Gate delays, shown in Fig.~\ref{fig1}c, include the four serial
junctions of JTL interconnect needed to connect gate to
gate. Latencies are no more than a few ps, but depend strongly on
initial state of the gate and the number of simultaneous
inputs. Latencies can be understood qualitatively is terms of
potential energy as shown in Fig.~\ref{fig1}d. In real designs,
intermediate pulse timings may occur that are neither well-separated
nor simultaneous. The curves in Fig.~\ref{fig1}c should be interpreted
as the best-case and worst-case timing extremes that envelope possible
gate delays.

Dual rail encoding solves the fundamental timing problems of inversion
in SFQ pulse-based logics. While global timing is stable as signals
travel through the wave pipeline at the rate determined by the ac
clock, significant local timing uncertainty arises from thermal noise,
parameter spread, and relative pulse arrival time. This translates to
latency in SFQ inverting logic due to race conditions. Dual rail
inversion is ``free'' in analogy to CMOS, and enables conventional
combinational logic. The cost of duplication in dual rail is partially
recouped when mapped onto our PCL gate primitives as the multiple
outputs are all available. The dual rail macros, shown in
Fig.~\ref{fig1}e, produce all the logical functions of a standard CMOS
gate library. The XOR3 macro implements a full adder with 70 JJs and
carry-out latency of 5\,ps corresponding to 12 levels of
two-input-gate logic per pipeline stage at 30\,GHz. Junction count is
about twice the number of transistors in a CMOS implementation
\cite{zhuang1992new}.

\begin{figure*}
\includegraphics[width=7.0in]{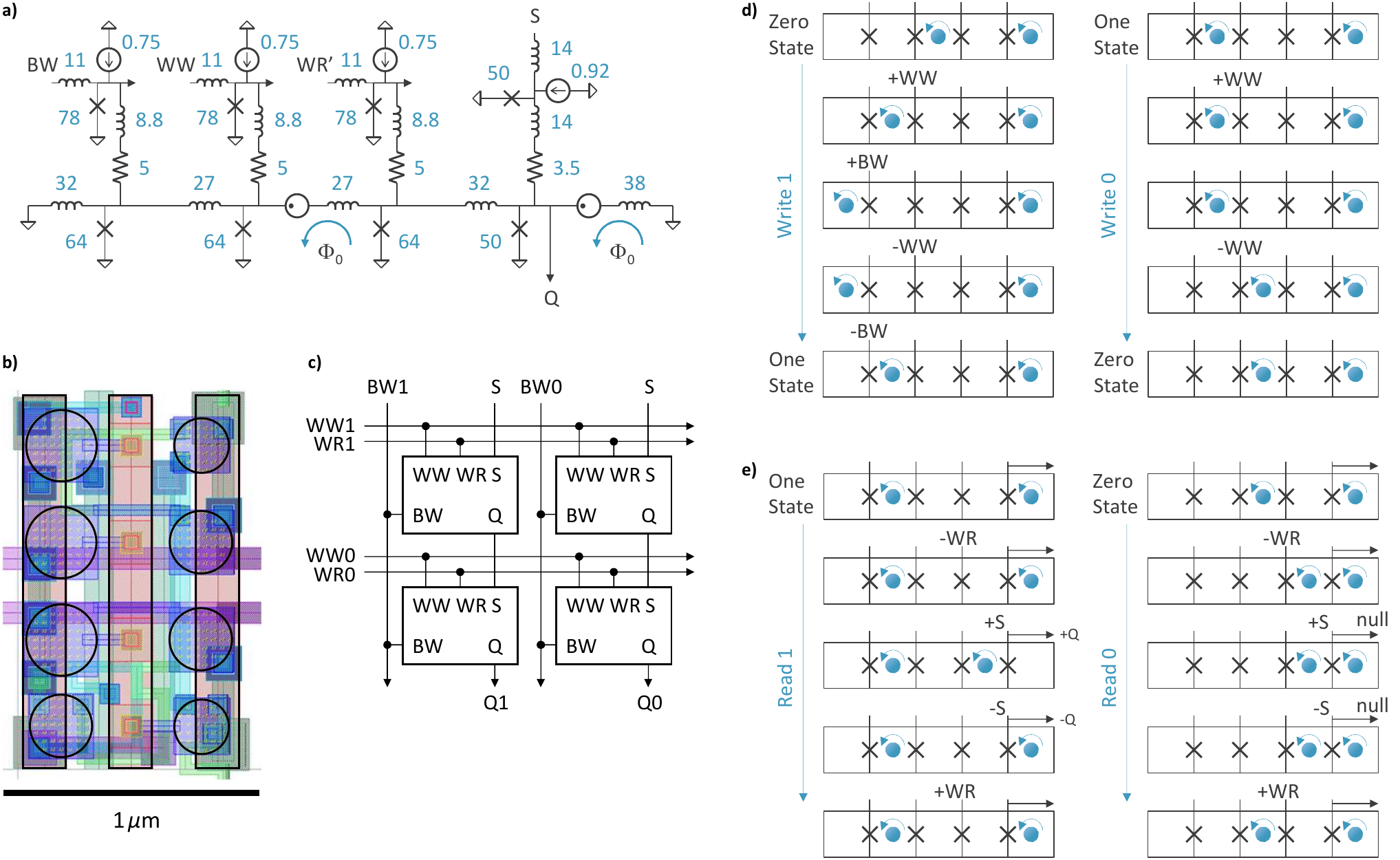}
\caption{\label{fig2} The JSRAM memory design. a) The unit cell
  schematic shows inductors in units of pH, JJ critical current in
  $\mu$A, peak AC current bias as a fraction of junction critical
  current, and resistors in $\Omega$. The schematic shows SFQ sources
  to initialize two storage loops, but this is accomplished with
  magnetic flux bias in the physical implementation. b) Physical
  layout of the unit cell has dimensions of 1$\times$2$\mu$m$^2$. High
  critical current density and high fill density of about 25\%
  accommodates the junctions (circular outlines). The flux bias is
  implemented in a single metal layer with a central primary and
  secondaries on each side (rectangular outlines). c) The memory
  fabric is a systolic array of unit cells. Word Write (WW), Bit Write
  (BW), and Word Read (WR) signals fanout to all cells in a vine
  configuration, whereas Strobe-to-Output (S-to-Q) Bit Read is a daisy
  chain through the array. d) Operation of the unit cell is
  illustrated as time sequences of initial state, event, final
  state. If the current of a stored SFQ interferes constructively with
  the current associated with an event in any JJ, the JJ critical
  current will be exceeded with the effect of moving the stored SFQ to
  the adjacent loop. Otherwise the event will be lost across the
  coupling resistor with no change in state. While Readout of state
  involves conditional propagation of the Bit Read, a cell that is not
  selected with Word Read will perform pass-through of the Bit Read
  irrespective of state.}
\end{figure*}

SFQ-compatible memory has historically demonstrated quite modest
density despite fundamental SFQ state retention in superconducting
loops. The difficulty in scaling superconducting memory centers on the
addressing, which requires fanout. Approaches can be categorized as:
active, which has the same signal levels and power distribution as
logic, i.e. Non-Destructive Read-Out (NDRO) \cite{dorojevets2001flux,
  burnett2018demonstration}; passive, meaning the array must be driven
with current levels at the perimeter, e.g.\ Vortex-Transitional
\cite{tahara1989experimental, nagasawa1995380, nagasawa1999high,
  semenov2019very}; and magnetic, which requires hysteretic tunnel
devices \cite{dayton2018experimental, nguyen2020cryogenic}. Passive
and magnetic approaches require drive levels corresponding to 100's of
SFQ pulses, which constrains density, power dissipation, throughput,
and latency. Drivers of current must be able to target and tune the
levels to accommodate analog addition of word and bit line signals in
the unit cell. The Vortex memory has transformers in the unit cell
that cannot be miniaturized below 0.25\,$\mu$m with practical drive
levels \cite{semenov2019very}. Magnetic memory devices would require
even larger currents to produce the demonstrated switching fields
\cite{gingrich2016controllable, birge2018ferromagnetic}.

Fig.~\ref{fig2}a shows the unit cell of a new Josephson memory, JSRAM,
an active array that can be understood as an optimized version of NDRO
with about 3$\times$ fewer JJs. The unit cell has four JJs for storage
and readout of state, and four AC-biased JJs that are JTL-based active
address lines. All signals and internal states are SFQ. Resistive
coupling to the address lines is used instead of the ``escape''
junctions used in RSFQ logic. This is feasible as the clock period is
long compared to the SFQ pulse width. The current scale in the LR
loops is set by the inductance, and the relaxation time is then set by
the resistance to small compared to the clock period. Simulated bias
current margins of the unit cell embedded in an array exceed $\pm$30\%
for global bias current amplitude.

Physical layout of the unit cell, shown in Fig.~\ref{fig2}b, achieves
densities for active memory even higher than contemplated for passive
arrays. The fabrication stackup proposed in a companion letter enables
4\,MB/cm$^2$ density, which represents a 600$\times$ increase in
density relative to legacy NDRO circuits implemented in a 250\,nm Nb
fabrication process.  Increased density of fabrication resources is
achieved across the stack including bias taps, junctions, and
inductive wiring.  Moving from 250\,nm to 50\,nm critical dimension in
the wiring layers increases inductance density by a factor of 25, and
moving to high-kinetic-inductance materials increases density by
another 10$\times$. The unit cell uses only the first four wiring
layers (M1-M4) for inductors, and a single metal layer to implement the
two SFQ flux biases, with mutual inductance of 0.5\,pH and 4\,mA
current in the primary. The top two layers are available for passive
transmission line wireup across subarrays.

The array, shown in Fig.~\ref{fig2}c is wave pipelined with throughput
at 30\,GHz, making it the functional equivalent of CMOS SRAM. Read and
Write operations are illustrated in Fig.~\ref{fig2}d. Important
features incorporated by the design have been published, including SFQ
readout on the bit line \cite{polonsky1995rapid}, and equal-energy
levels for the different flux states in the memory cell based on
moving an SFQ between superconducting loops \cite{nair2018ternary,
  toomey2019bridging}.

\begin{table}
\caption{\label{table1}Comparison of our proposed SFQ technology to a
  tensor processor unit (TPU) implemented in leading-node CMOS.}
\begin{ruledtabular}
\begin{tabular}{lccc}
 & SFQ & TPUv4i & Ratio \\
\hline
Process node          & 28\,nm & 7\,nm & 4 \\
MAC area ($\mu$m$^2$) & 6,700\footnote{Using a derated 50\% critical resource utilization relative to JSRAM.} &
       600\footnote{An estimated 10\% of the die area dedicated to MAC units.} & 11 \\
Clock (GHz)           & 30 & 1.05  & 29 \\
Computational density (M\,op/s/$\mu$m$^2$) & 4.5 & 1.75 & 2.6 \\
Energy efficiency (op/pJ)      &
       69\footnote{Including a 325\,W/W cryocooling overhead.} & 3.1 & 22 \\
\end{tabular}
\end{ruledtabular}
\end{table}

Table~\ref{table1} compares the size and performance of a PCL bfloat16
multiply-accumulate unit (MAC) to the TPUv4 MAC implemented in
leading-node \cite{jouppi2021ten}. The BF16 multiplier contains an
8$\times$8 array multiplier for the significand and an 8-bit adder for
the exponent, with a small additional overhead for alignment and
normalization. The total device count for PCL bf16 MAC is 14,330\,JJs,
including the systolic array 8$\times$8 PCL multiplier and the
single-precision floating-point 2-bit accumulator. The multiplier
contains 49 and2 gates, 7 half-adders, and 42 full-adders for a total
of of 4,284\,JJs. The single-precision floating-point 32-bit
accumulator has 4,163\,JJs. The area of the PCL MAC has been estimated
by derating JSRAM design density by a factor of two. As shown in the
table, power efficiency at the level of on-chip logic is 20$\times$
better than CMOS while including a 325$\times$ cooling
overhead. Computational density of SFQ at 28\,nm is higher than 7\,nm
CMOS, in part due the high clock rate of 30\,GHz and in part due to
the high utilization of chip real-estate used for logic.

PCL and JSRAM technology enable even higher energy efficiency on the
system level, up to 100$\times$ due to superconducting
interconnect. Low power dissipation of the JSRAM enables efficient
chip stacking providing sufficent SRAM-to-compute. Ultimate system
performance is determined by capacity and bandiwdth to cryo-DRAM that
is an area of active research \cite{ware2017superconducting,
  wang2018dram, bae2019characterization, lee2019cryogenic,
  lee2021cryoguard, garzon2021gain}.

In conclusion, we have put forward advances in SFQ logic and memory
design that are essential to scaling. AC-biased PCL logic provides
economical dual-rail macros with low-latency inversion, solving one of
the central problems of pulse driven logics. JSRAM is a
superconducting memory providing SRAM capability and high
density. Both technologies are enabled by the high-integration-density
fabrication stack described in a companion letter. PCL logic projects
to 12 levels of logic per stage with a 30\,GHz clock, and a
computational density on par with 7\,nm CMOS. At the gate level,
energy efficiency is 20$\times$ better than CMOS including cryocooler
efficiency. JSRAM memory projects to 4\,MB/cm$^2$ with throughput at
the 30\,GHz clock rate and a latency of 1\,ps per stage in the active
array. These technologies enable practical architectures with
100$\times$ better power efficiency at the system level and 30$\times$
higher clock rate.

\begin{acknowledgments}
Work at imec and imec USA is supported by imec INVEST+ and by Osceola
County.
\end{acknowledgments}

\subsection*{Author Contributions}
All authors contributed to the work equally.

\nocite{}
\bibliography{compendium}

\end{document}